# A Primer on Generative AI for Telecom: From Theory to Practice

Xingqin Lin, Lopamudra Kundu, Chris Dick, Maria Amparo Canaveras Galdon, Janaki Vamaraju, Swastika Dutta, Vinay Raman

NVIDIA

Email: {xingqinl, lkundu, cdick, acanaveras, jvamaraju, swastikad, vraman}@nvidia.com

*Abstract*—The rise of generative artificial intelligence (GenAI) is transforming the telecom industry. GenAI models, particularly large language models (LLMs), have emerged as powerful tools capable of driving innovation, improving efficiency, and delivering superior customer services in telecom. This paper provides an overview of GenAI for telecom from theory to practice. We review GenAI models and discuss their practical applications in telecom. Furthermore, we describe the key technology enablers and best practices for applying GenAI to telecom effectively. We highlight the importance of retrieval augmented generation (RAG) in connecting LLMs to telecom domain specific data sources to enhance the accuracy of the LLMs' responses. We present a real-world use case on RAG-based chatbot that can answer open radio access network (O-RAN) specific questions. The demonstration of the chatbot to the O-RAN Alliance has triggered immense interest in the industry. We have made the O-RAN RAG chatbot publicly accessible on GitHub.

## I. INTRODUCTION

The rise of generative artificial intelligence (GenAI) is propelled by several converging forces, including the powerful transformer architecture, advanced deep learning algorithms, vast amounts of data, and high-performance computing accelerated by graphics processing units (GPUs). The technological advancement has led to the creation of powerful GenAI models, particularly the large language models (LLMs) such as generative pre-trained transformer (GPT). The exceptional performance of the GenAI models (e.g., Open AI's GPT-4o) and their access through user friendly interfaces have brought text and image generation to the forefront of daily and commonplace conversations. Now GenAI is transforming the world, driving innovations in a wide range of industries and emerging applications.

The telecom industry is among the firsts to embrace GenAI in a variety of use cases such as customer service and network operations [1]. Chatbots powered by GenAI can assist customers with their queries, troubleshoot technical issues, and provide personalized recommendations for telecom services and products. By automating network operation tasks such as network optimization and predictive maintenance, GenAI is helping telecom industry in reducing operational costs and improving network efficiency. Overall, the potential of GenAI for telecom is vast and will continue to grow as the technology evolves [2].

Inspired by the great potential of GenAI, recently there have been a number of research works applying a multitude of GenAI models to the telecom domain. There is a growing trend of applying large GenAI models to telecom. An overview of large artificial intelligence (AI) models for the sixth-generation (6G) wireless networks is presented in [3], focusing on opportunities, challenges, and research directions. The work [4] explores how large GenAI models can be used for designing, configuring, and operating wireless networks. LLMs are leveraged by the work in [5] to develop a framework that orchestrates edge AI models. Besides, the work in [6] examines the current capabilities and limitations of LLMs and discusses the use cases of LLMs for telecom as well as the research directions. However, GenAI for telecom is still a largely uncharted territory for standards development of next generation wireless networks [7].

While the existing work has explored the potential of GenAI for telecom, there remains a gap between the research outcomes and their real-world applications. The existing literature primarily focuses on the theory or vision of GenAI for telecom, often overlooking the implications and challenges that exist in practice. This article aims to address this gap by bridging between the intricacies of theories and their manifestation into use case enablement in practice. To that end, we first examine the commonly used GenAI models for telecom by highlighting their theoretical foundations and relevance to key use cases. Then we specifically focus on LLMs and provide an overview of the practical applications of LLMs as can be found in the telecom industry today. Furthermore, we describe the key technology enablers and best practices for applying GenAI to telecom efficiently and effectively. In particular, we highlight a popular technique – retrieval augmented generation (RAG) – that can link LLMs to domain specific and/or enterprise data sources to improve accuracy, increase user trust, and reduce hallucinations [8]. We illustrate the power of RAG-augmented LLMs for open radio access network (O-RAN) standards. With its knowledgebase augmented with domain specific information from the technical specifications published by the O-RAN Alliance, the chatbot can succinctly answer any question or query related to O-RAN specification in natural language. We also made the O-RAN RAG chatbot open sourced and publicly accessible on GitHub [9]. This is the first-of-its-kind, GenAI powered, and RAG augmented open RAN standards chatbot that is made freely available to the global telecom ecosystem. After we presented the chatbot to the O-RAN Alliance and made it publicly accessible on GitHub, there have been other works along this line, e.g., the recent work in [10] presents Telco-RAG for processing the 3rd generation partnership project (3GPP) specifications.

## II. PRELIMINARIES OF GENERATIVE AI MODELS FOR TELECOM

GenAI is used to produce new but similar samples distributed according to some unknown distribution of the existing samples. The goal of GenAI modeling is to develop a model that learns the unknown distribution so that we can use it for sampling. A multitude of GenAI models have been applied to telecommunications problems, including variational autoencoder (VAE), generative adversarial network (GAN), normalizing flow (NF), diffusion model (DF), and transformer-based models. In this section, we present the preliminaries of these GenAI models, as illustrated in Figure 1.

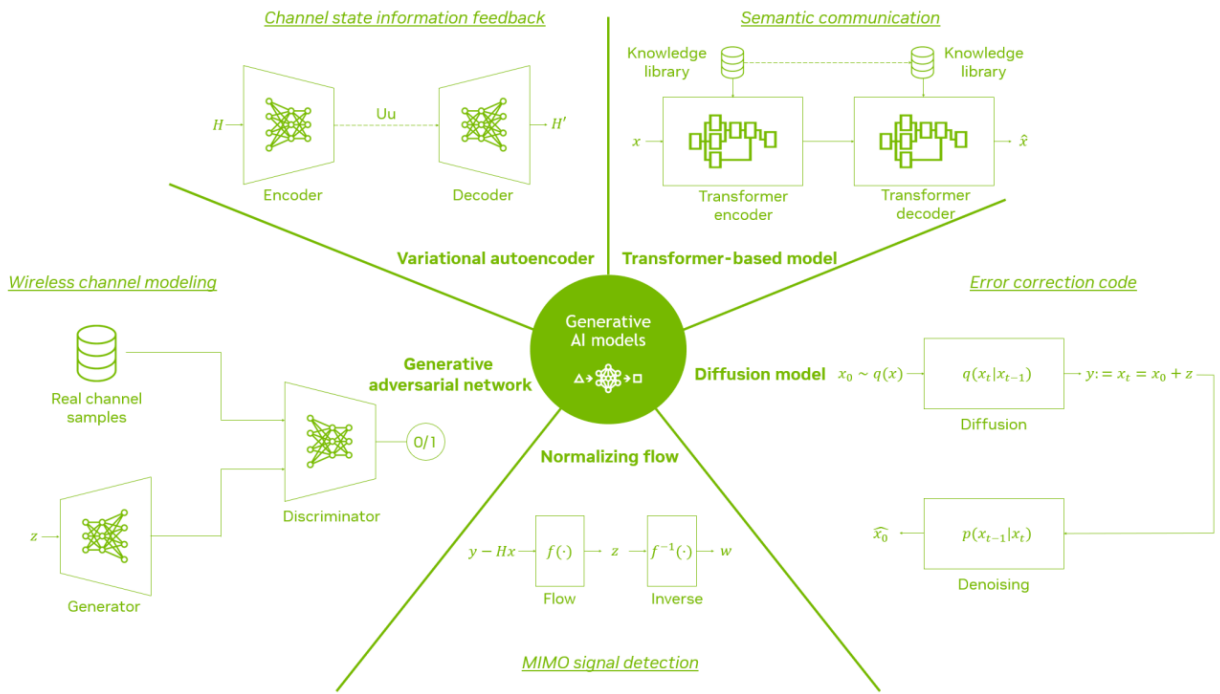

**Figure 1: An overview of GenAI models and their example use cases in communications.**

*Variational Autoencoders*: A VAE consists of a probabilistic encoder and a probabilistic decoder. The encoder maps an input to a distribution over a latent space, whereas the decoder maps a point sampled from the distribution over the latent space to a distribution over the input space from which we can sample. Neural networks can be trained to express the encoder and decoder functions to yield a performant encoding-decoding scheme. One potential use case of VAEs in telecommunications is their application to channel state information (CSI) feedback in massive multiple-input multiple-output (MIMO) systems, where CSI can be encoded at user equipment (UE) and decoded at the serving base station [11].

*Generative Adversarial Networks*: A GAN consists of a generator and a discriminator. The generative network takes a random input to generate data with the same distribution as the training data, while the discriminative network takes an input that can be either "true" or "generated" and returns as output the probability of the input to be a "true" one. After training, the generator becomes capable of producing realistic data and the discriminator learns the characteristics of the true data. One potential use case of GANs in telecommunications is their use in wireless channel modeling [12]. Specifically, real channel data can be used to train a channel data generator and a channel data discriminator, and then the resulting channel data generator can be used to generate synthetic channel data for the target radio propagation scenario.

*Normalizing Flows*: An NF uses a sequence of invertible and differentiable mappings for transformation between a simple distribution such as a Gaussian distribution and a complex target distribution. In the generative direction, one can sample a point from the simple distribution and push it forward in the mappings to a point in the target distribution. Conversely, the normalizing direction is a pushforward from the target distribution to the simple distribution. One potential use case of NFs in telecommunications is their use in MIMO signal detection, where the noise distribution can be approximated through an NF [13].

*Diffusion Models*: A DF is composed of a forward diffusion process and a reverse diffusion process. In the forward diffusion process, an input sample is progressively combined with Gaussian noise over a sequence of steps, resulting in an output of Gaussian noise when the number of steps becomes large. In the reverse diffusion process, a model is trained by removing the noise step by step to maximize the likelihood of the training data. One potential use case of DFs in telecommunications is their use in error correction code, where the forward channel corruption can be modeled as a series of diffusion steps that can be reversed iteratively at the receiver [14].

*Transformer-based models*: A transformer mainly comprises of tokenizers, embedding layers, and transformer layers. The major idea behind the transformers is called attention or self-attention that enables detection of the subtle relationships of sequential data even when the data elements are distant. A self-attention layer encodes each input entity with the global contextual information from the complete input sequence. Furthermore, multi-head attention uses multiple self-attention blocks to capture multiple relationships in the input sequence, allowing for parallel processing and scalability to highly complex models and large data sets. One potential use case of transformers in telecommunications is their use in semantic communication, where transformers can be employed to extract abstract semantic information at the transmitter and reconstruct information at the receiver [15].

The different types of GenAI models have varying levels of performance in terms of the quality of generation outputs, the diversity of mode coverage, and the speed of sampling. Combining the advantages of the GenAI models, when possible, can create more powerful GenAI models for telecom. While GenAI models have shown great potential for enabling various emerging telecom use cases in simulated or lab environments, transformer-based LLMs are among the most popular GenAI models that are already finding practical applications in the current telecom industry. These real-world applications of LLMs in the telecom domain are detailed in the next section.

## III. LLMs FOR TELECOM

Harnessing the power of LLMs, the telecom industry is leaping forward to transform its business, all the way from improving the network's operational efficiency to reshaping the customer experience and employee productivity.

*Customer Support*: Domain-specific LLMs are increasingly gaining popularity in telecom's customer care services and related applications. With LLM based customer service platforms, telecom companies can 1) promptly troubleshoot billing issues like inaccurate charges, service failures or missing discounts, 2) curate resolution for each issue based on customer interaction history and context, 3) timely communicate the proposed resolution to the customer using natural language generation, and 4) follow up with the customer to ensure satisfaction and offer additional services, promotions or discounts, leveraging sentiment analysis and natural language interaction prompts. As one example, Amdocs has recently announced a purpose-built GenAI platform for telecom called "amAIz" that fine-tunes generic LLMs with telecom-specific data to specialize in customer engagement. The platform utilizes LLMs trained on customer bills, orders, business policies, processes, and historical agent transcripts to provide automated billing service using natural language generation.

With the help of customized LLM-based agents, customer service assurance teams can now decipher complex network logs translated into natural language and distill complex information into concise incident summaries to automate customer communications and speed up time-to-resolution. As an example, ServiceNow has integrated GenAI capabilities into its "Now Assist" platform, with features powered by fine-tuned LLMs, to enable use cases like automated service assurance.

*Field Technician Assistance*: LLM powered telecom chatbot applications are assisting network technicians in the field by promptly diagnosing critical network issues and recommending efficient resolutions by synthesizing large volumes of technical manuals. Such LLM-based digital assistants greatly improve technician's efficiency at the cell-site, resulting in faster query resolution, enhanced troubleshooting accuracy, and minimized back-office workload (e.g., reduced frequency of supervisor calls). As one example, GenAI powered platform "Baioniq" from Quantiphi offers customized LLM-based virtual assistants as copilots for field technician assistance.

*Network Diagnostics, Management, and Planning*: GenAI enables telecom professionals to have dynamic conversations with network data in natural language by the virtue of LLMs. LLMs can help analyze the network logs, visualize network service areas with performance issues, and recommend deployment strategy of technicians in the field. GenAI is also being used to proactively monitor network performance, predict possible network degradations by analyzing logs and telemetry along with historical datapoints, and provide new insights for repair planning in plain language. As one example, Kinetica has recently unveiled "SQL-GPT" which leverages LLM and vectorized processing to enable telecom professionals to have interactive dialogue with network using natural language and through visualization of datapoints on a map.

Furthermore, blending intelligence with automation, GenAI can help telcos with recommendation of network design and configurations, including placement of cell towers. As one example, TCS's digital twin platform "TwinX" is harnessing LLM-enabled insights to provide data analytics for network planning.

*Telecom Standards Chatbots:* Telecom standards play a crucial role in network deployment, where the right understanding of standards is essential all the way from making interoperable and standards-compliant network equipment products to plugging them together to build the network infrastructure and establishing wireless connectivity between the base stations and UE. Understanding nitty-gritty details of standards spread across a slew of complex documents and their interrelations can be quite challenging and would require domain expertise. LLM-based GenAI augmented with advanced features like RAG can simplify and streamline access to complex specifications, enhancing collaboration and understanding of industry standards. As one example, the O-RAN chatbot open-sourced by NVIDIA is a RAG enabled LLM platform that provides quick summary of queries related to O-RAN standards, all the way from simple, one-liner questions and answers (Q&As) to complex summarization questions requiring consultation from multiple standards documents across various O-RAN working groups. Such tools make highly specialized telecom technical specifications accessible to broader ecosystem partners (especially academia and niche industry players) who may not be directly involved in the discussion and development of the telecom standards.

Overall, the global telecom ecosystem is actively aiding the industry's march towards fully embracing the power of GenAI, with a growing number of LLM-enabled solutions unlocking new levels of insight and efficiency into network lifecycle, all the way from the planning phase to deployment, operations, and maintenance. Unleashing the full potential of LLM to enable these use cases would require an understanding of the key design principles behind building and deploying LLM-based solutions specifically for telecom, which is addressed in the following section.

## IV. DESIGN ASPECTS

Cloud-based services utilizing GenAI foundation models, such as general-purpose LLMs, offer a swift entry point into the realm of GenAI technology for telecom. Yet, these services typically cater to a wide array of functions and lack in the richness of training on telecom specific datasets, thus constraining their efficacy for specific telecom needs. In this section, we discuss key design aspects of GenAI development and deployment for building customized AI applications that are tailored to telecom specific requirements, as illustrated in Figure 2.

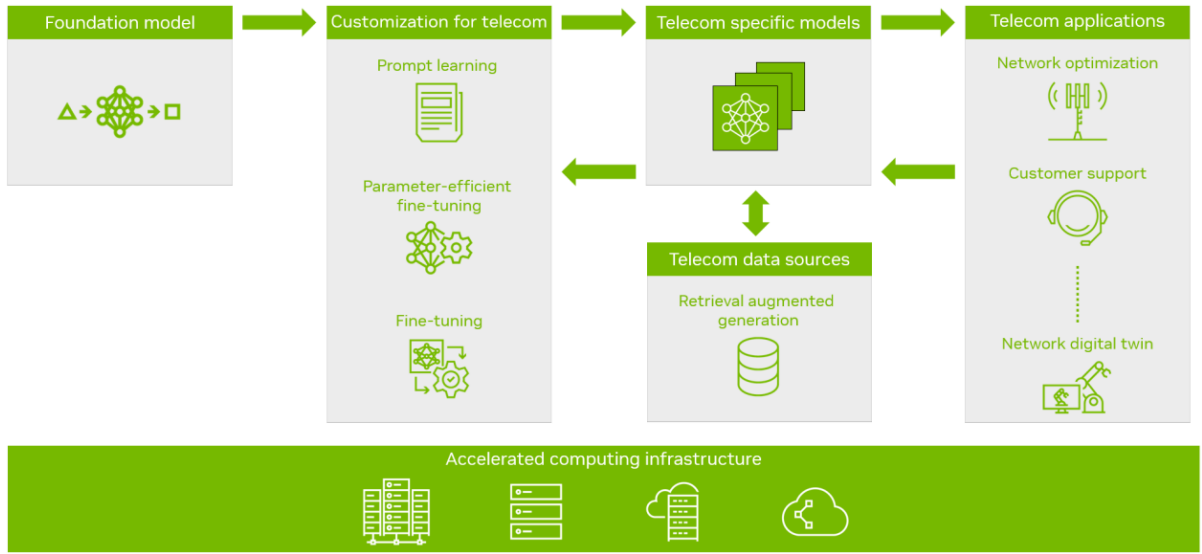

**Figure 2: An overview of the life cycle of building GenAI applications for telecom.**

*A. Accelerated Compute*

The backbone of LLMs is the transformer. Transformer models are among the largest neural network models, with GPT-4 estimated to employ approximately 1.76 trillion parameters. The recent trend shows that the compute requirements for transformer AI models are growing at a rate of 275-fold every 2 years, in contrast to the more modest 8-fold increase for other AI models. Training these enormous models in a reasonable amount of time requires both optimized silicon and multi-GPU compute infrastructure. The state-of-the-art GPUs like Hopper and Ada Lovelace provide critical optimizations to reduce training time, lower memory requirements, and build efficient scale-out of AI super computers, offering proportional increase in compute capability as the number of GPUs grows in the cluster.

For many computing devices, the amount of compute scales with the data type bit precision. For example, for floating point (FP) data types like FP64, FP32, and FP16, the Hopper H100 SXM5 GPU delivers 33.5, 66.9, and 133.8 tera floating point operations per second (TFLOPs) respectively–a doubling of compute capability for each halving of the bit precision. However, low-precision data types may negatively impact model performance. Furthermore, certain layers in a neural network may be more sensitive to data type precision than others. Therefore, it is common to employ a mixed-precision approach in the neural network architecture.

The Hopper and Ada Lovelace GPUs incorporate a transformer engine that includes a combination of software and custom tensor cores to support mixed precision. The transformer engine introduces a low-precision FP8 data type that enables a mixture of FP16 and FP8 precisions to dramatically accelerate AI training and inference. To preserve model accuracy when using the low-precision FP8, a novel auto-scaling strategy is employed, wherein the transformer engine dynamically selects between FP8 and FP16 calculations and automatically handles the re-casting and scaling between FP8 and FP16 in each layer. Since FP8 is not a native data type for machine learning frameworks today (such as PyTorch and JAX), the transformer engine offers a C++ based application programming interface (API) through its software component that integrates with popular LLM libraries.

Accelerated compute capability is crucial for telecom LLMs to facilitate efficient handling of large telecom datasets and provide real-time responses. For example, processing vast amounts of data continually generated from network operations and service usage requires substantial computational power. Enabling that while managing network operational cost and maintaining sustainability is only possible with accelerated compute platforms. GPU-based accelerated compute platforms are inherently designed to handle large workloads through efficient parallel processing, thereby consuming less power compared to traditional central processing units (CPUs) for the same amount of work. Besides, real-time analysis of such enormous amount of data in telecom is warranted for timely decision-making in LLM applications like customer chatbots, field technician assistance or network diagnostics and repair planning. Accelerated compute with low-latency inference capability can provide quick responses for such applications, improving user experience, operational efficiency and service quality. As telecom companies will embrace more and more LLM-driven services in the foreseeable future, accelerated compute will enable them to scale their compute needs efficiently and develop more innovative services without being constrained by compute limitations.

*B. Information Retrieval and Customization*

There are foundation LLMs meticulously pre-trained on large datasets by harnessing the power of accelerated computing. However, to develop effective LLMs for telecom, it is critical that the LLMs incorporate telecom domain knowledge and/or enterprise data sources. This can be done with RAG which is able to connect LLM prompts with the information retrieved from external sources, leading to improved accuracy and reliability of the LLMs.

Figure 3 shows a typical RAG pipeline, consisting of offline document ingestion and online query and response generation with the following steps:

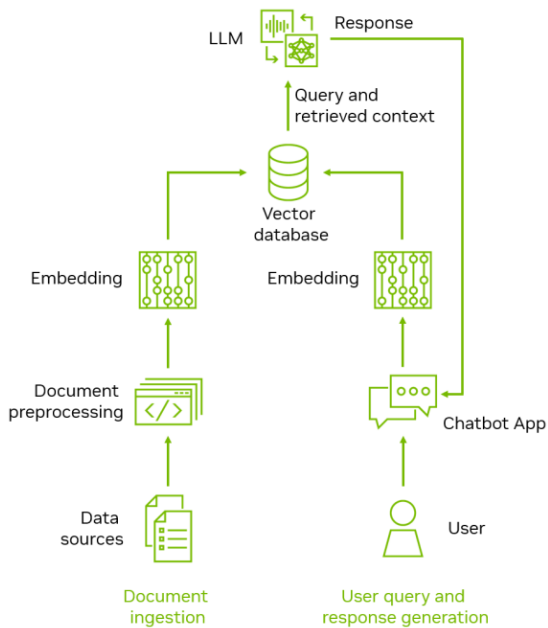

**Figure 3: An Illustration of RAG pipeline with document ingestion and query flows.**

- Step 1 (*Documents ingestion*): To begin with, the target data sources are loaded into the RAG pipeline.
- Step 2 (*Documents preprocessing*): Next, the loaded documents are transformed, e.g., splitting long text into shorter segments to fit within the subsequent embedding model.
- Step 3 (*Embeddings generation and storage*): In this step, embeddings are generated for the documents to represent the data as numerical vectors, followed by storing the generated embeddings in vector databases.
- Step 4 (*Query and context retrieval*): When a user query is submitted, the query is first embedded and then compared with the stored vectors in the vector databases to retrieve contextual information.
- Step 5 (*Response generation*): Based on the user query and the retrieved contextual information, the LLMs generate appropriate responses.

In addition to RAG, other popular LLM customization techniques include prompt engineering and learning, parameter-efficient fine-tuning (PEFT) that selectively adds or updates a few parameters or layers of the original LLM architecture, and general fine-tuning methods, e.g., supervised fine-tuning and reinforcement learning with human feedback (RLHF). Fine-tuning techniques like PEFT or RLHF can be powerful tools to customize a pretrained LLM for telecom by updating the parameters of the LLM based on telecom domain knowledge and/or enterprise data sources. Compared to the RAG based customization, fine-tuning is more resource intensive but yields higher accuracy for certain use cases. However, RAG and fine-tuning are not mutually exclusive technologies, but rather can be used in tandem. When starting to build a telecom application with an LLM, RAG can be firstly applied to quickly improve accuracy. If the application demands even higher accuracy, fine-tuning can be further used to customize the LLM. In a nutshell, the best approach of customizing an LLM for telecom depends on the application's specific requirements by considering resource availability and computational constraints.

RAG-enhanced LLMs can be very effective and useful for enabling telecom specific use cases like network planning or telecom standards chatbot. Take telecom standards chatbot for example. Generic LLM models can be enhanced with telecom-specific standards documents through RAG mechanism, so that the LLM models become more accurate and efficient in answering domain-specific questions related to telecom standards. A case study illustrating the efficacy of this example is detailed in the following section.

## V. CASE STUDY: O-RAN RAG CHATBOT

In this section, we present a case study on the use of the RAG techniques to connect LLMs to telecom domain knowledge contained in a few hundreds of specification documents from the O-RAN Alliance.

### A. Design of Core Functional Components

To start with, we describe the core functional components that underpin the performance and reliability of the O-RAN RAG chatbot, specifically tailored to efficiently manage and utilize vast amounts of data.

*Document Ingestion and Preprocessing:* Data chunking and filtering are important to effectively use RAG with large datasets. Data chunking involves dividing a large corpus of texts into smaller, more manageable chunks. Filtering complements chunking by going through these chunks to remove irrelevant, redundant, or low-quality information. In our tailored implementation for the O-RAN specification documents, we conducted experiments with various chunk sizes to determine the most effective configuration for the RAG pipeline. We ultimately selected an optimal chunk size of 500 words, with an overlap of 100 words between chunks. This configuration was chosen to ensure continuity and context retention across different sections of the documents, enhancing the coherence of retrieved information. Post-chunking, we applied a series of targeted filters to refine the quality of the text passages. Firstly, we removed all non-English characters from passages to maintain consistency and avoid encoding errors during further data processing. Secondly, we removed non-essential separation characters such as break lines to create a smoother, more continuous flow of text. Thirdly, we discarded overly short chunks because they often needed more contextual information for meaningful embeddings.

*Embedding Generation and Storage:* Embedding models employ transformer encoder blocks, which process input tokens (such as questions or passages) to generate a corresponding vector or embedding. We leveraged NVIDIA NeMo retriever in this work. Specifically, we used the NV-Embed-QA-Mistral-7B model as the embedding model and the ms-marco-MiniLM-L-6-v2 reranking model to optimize the order of the results for relevancy. We adopted the GPU-accelerated Facebook AI similarity search (Faiss) library as the vector database for storing and retrieving these embeddings. Faiss is designed for efficient and rapid search of similar vectors, facilitating quick retrieval of relevant information from the database.

*Query and Context Retrieval:* Prompt enhancements heavily influence the quality of the answers generated by an LLM. An augmented prompt in a RAG system includes several components, including instruction, context, input data, and output indicator. In this work, we exploited several considerations to create an effective prompt, including clearly defining the bot's role, introducing bot tone phases, incorporating default answers, and adding instructions.

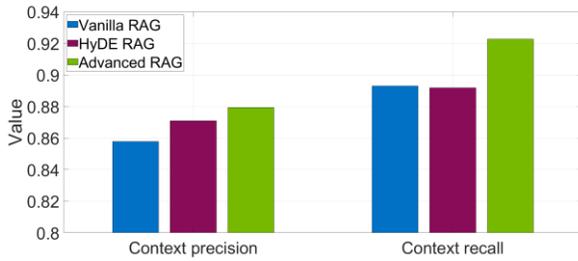

Figure 4: Information retrieval accuracy comparison of Vanilla RAG, HyDE RAG, and Advanced RAG.

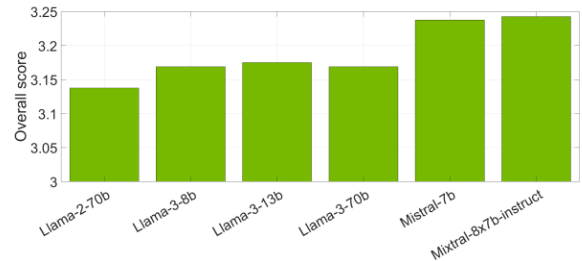

Figure 5: End-to-end testing with LLM-as-a-judge (GPT-4) rating on different LLMs using Advanced RAG.

*Response Generation:* Fact-checking guardrails ensure the accuracy and reliability of the responses generated by an LLM. This verification step confirms that the responses provided by the LLM are based on the information extracted from the augmented prompt and are not solely dependent on the model's training data. In this work, we implemented fact-checking using the NVIDIA NeMo Guardrails library. This open-source toolkit facilitates the integration of programmable guardrails into LLM-based conversational systems, enforcing specific output controls that adhere to predefined content restrictions or guidelines.

We also adopted LangChain framework to integrate various components of the chatbot, manage the flow of data, and ensure that each component interacts seamlessly with the others. Finally, we leveraged Streamlit user interface as the front-end through which users can interact with the chatbot.

### B. Experiments and Evaluation Results

The previous subsection describes the basic setup of the O-RAN RAG chatbot, refereed hereafter as "Vanilla RAG". One limitation of the Vanilla RAG is its inaccuracy or incompleteness in answering certain questions. This is mainly caused by incomplete retrieval of information from the vector database. To overcome the drawback, we need to evaluate other potential retrieval optimization techniques and LLMs.

To enhance the completeness of the retrieved information beyond Vanilla RAG, we tested two additional types of retrieval optimization techniques termed as "Hypothetical document embeddings (HyDE) RAG" and "Advanced RAG" which aim to improve the quality and quantity of documents fetched by the retriever.

*HyDE RAG* uses an LLM to generate a hypothetical answer to the user query and then performs a similarity search with both the original question and the hypothetical answer, instead of doing the similarity search with just the user's original query. After retrieving a set of documents using similarity search, we re-rank the documents with the ms-marco-MiniLM-L-6-v2 cross-encoder model, and pass the top ten documents to the LLM for response generation. This technique outperforms standard retrievers and eliminates the need for custom embedding algorithms, but it can occasionally lead to incorrect results as it is dependent on another LLM for additional context. In this work, we used Llama-2-70b to generate the hypothetical answer.

*Advanced RAG* utilizes query transformation which changes the question fed into the RAG pipeline to find relevant chunks based on more than the user's original question. Specifically, we implemented a query transformation technique that uses an LLM to generate multiple sub-queries from a single user input. Using the Llama-2-70b model, the RAG pipeline first generates ten similar questions to the user's original query. For each of these queries, the relevant documents are retrieved from the vector database. Following deduplication (i.e., eliminating duplicate copies of the retrieved documents), the documents are filtered and re-ranked using the ms-marco-MiniLM-L-6-v2 cross-encoder model, and the top ten documents are passed as the context to the LLM for response generation. The query transformation enables a more thorough and nuanced comprehension of complex topics typically scattered across various documents, enhancing the chatbot's capability to deliver relevant and accurate information.

In Figure 4, we compare the information retrieval accuracy of Vanilla RAG, HyDE RAG, and Advanced RAG in terms of context precision and context recall by using the Ragas framework. The context precision metric evaluates whether the items in the contexts that are relevant to the ground truth are ranked higher than irrelevant ones. The context recall metric measures the extent to which the retrieved context aligns with the ground truth. Both metrics have values ranging between 0 and 1, with higher values indicating better performance. Figure 4 shows that HyDE RAG has a higher context precision value than Vanilla RAG, while Vanilla RAG has a slightly higher context recall value than HyDE RAG. Meanwhile, Advanced RAG outperforms both Vanilla RAG and HyDE RAG in terms of both context precision and context recall.

End users are more interested in the quality of chatbot responses than the retrieval performance. However, judging the quality of chatbot responses is challenging, as the requirements are usually broad and loosely defined. Recently, LLM-as-a-judge has been increasingly utilized by the LLM community, with many using GPT-4 to evaluate their LLM outputs. In Figure 5, we also adopt LLM-as-a-judge (GPT-4) to evaluate the quality of the O-RAN RAG chatbot responses under different open-source LLMs augmented by Advanced RAG, including Llama-2-70b, Llama-3-8b, Llama-3-13b, Llama-3-70b, Mistral-7b, and Mixtral-8x7b-instruct. We can see that the scores of the three Llama-3 models are quite close, and they outperform Llama-2-70b. Besides, Mistral-7b and Mixtral-8x7b-instruct have higher scores than the Llama models, with Mixtral-8x7b-instruct rated slightly higher than Mistral-7b.

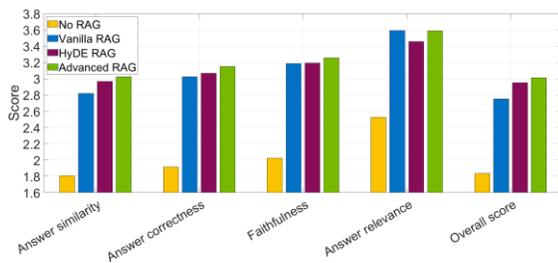

**Figure 6: End-to-end testing with LLM-as-a-judge (GPT-4) rating on Mixtral-8x7b-instruct with No RAG, Vanilla RAG, HyDE RAG, and Advanced RAG.**

Finally, we compare the quality of chatbot responses by using LLM-as-a-judge (GPT-4) to rate Mixtral-8x7b-instruct with No RAG, Vanilla RAG, HyDE RAG, and Advanced RAG. In Figure 6, we consider evaluations along five dimensions: answer similarity, answer correctness, faithfulness, answer relevance, and overall score considering all the four metrics. The answer similarity metric measures the semantic similarity between the generated response and the ground truth. The answer correctness metric assesses the accuracy of the generated response compared to the ground truth. The faithfulness metric gauges the factual consistency of the generated response against the retrieved context. The answer relevance metric measures the pertinency of the generated response with respect to the given prompt. Figure 6 shows that even Vanilla RAG significantly outperforms No RAG in terms of all the metrics. It is also shown that HyDE RAG has similar or higher scores than Vanilla RAG in all the metrics except the answer relevance metric. Besides, Advanced RAG performs the best consistently across all the metrics than the other schemes. In summary, Figure 6 illustrates the effectiveness of different RAG types in augmenting the LLM with domain specific information from the technical specifications published by the O-RAN Alliance.

## VI. CONCLUSIONS AND FUTURE OUTLOOK

The rise of GenAI has the potential to transform the telecom industry by driving innovation, improving efficiency, and delivering superior customer services. This paper has provided a primer on GenAI for telecom by overviewing the GenAI models and their relevance in telecom, examining practical LLM applications for telecom, and describing key technology enablers needed for effectively applying GenAI to telecom. In particular, we have highlighted the importance of RAG in combining the capabilities of LLMs with telecom domain specific and/or enterprise data sources. Furthermore, we have presented an O-RAN RAG chatbot to demonstrate the power of RAG-augmented LLMs for telecom and made the chatbot publicly accessible.

We conclude by pointing out some of the important topics for future work.

*Telecom domain specific datasets*: Specifications developed by standards bodies such as the 3GPP and O-RAN alliance provide valuable datasets for customizing foundation LLMs in telecom applications. However, there is an urgent need for more telecom knowledge datasets to be made publicly available to facilitate further advancement of GenAI for telecom.

*Multi-modal GenAI for telecom*: Foundation LLMs have been mostly trained on vast amounts of text data. In telecom, there exist a multitude of different data modalities such as radio signals and 2D/3D environment data from camera, radar, and LiDAR. The integration of multi-modal capabilities into foundation GenAI models deserves further exploration.

*Standardization aspects:* Though standards bodies such as the 3GPP and O-RAN Alliance have started to embrace AI, GenAI is still a largely uncharted territory for telecom standards. Standardization efforts may focus on interoperability and compatibility between different GenAI models and platforms used in the telecom industry.

## BIOGRAPHIES

**Xingqin Lin** is a Senior 3GPP Standards Engineer at NVIDIA.

**Lopamudra Kundu** is a Senior Standards Engineer at NVIDIA.

**Chris Dick** is a Wireless Architect at NVIDIA.

**Maria Amparo Canaveras Galdon** is a Senior Solutions Architect at NVIDIA.

**Janaki Vamaraju**, is a Staff AI Architect at NVIDIA.

**Swastika Dutta** is an AI Solutions Architect at NVIDIA.

**Vinay Raman** is a Senior Deep Learning Scientist at NVIDIA.